\documentstyle[prd,aps]{revtex}
\begin{document}
\draft
\twocolumn[\hsize\textwidth\columnwidth\hsize\csname@twocolumnfalse\endcsname
\title{Dynamically Broken Scale Invariance and Cosmology}
\author{Shih-Yuin Lin~\cite{lin}}
\address{Department of ElectroPhysics, Chiao Tung University, Hsin Chu, Taiwan 
300, R.O.C.}
\author{Kin-Wang Ng~\cite{ng}}
\address{Institute of Physics, Academia Sinica, Taipei, Taiwan 115, R.O.C.}
\date{September 1996}
\maketitle

\begin{abstract}
A new scalar-tensor theory of gravity induced by dynamically broken scale 
invariance is proposed, and its cosmological implications are discussed. 
It is found that the model admits an inflation via the Hawking-Moss bubbling, 
but the inflation rate remains undetermined due to the strong gravity limit. 
In light of this, scale-invariant metric perturbations having a dominant 
tensor component can be generated without slow-rollover. In addition, 
the deviation from the standard hot big-bang is vanishingly small after 
inflation.
\end{abstract}

\pacs{PACS numbers: 04.50.+h, 98.80.Cq}
\vspace{2pc}]
                                                                                
Einstein gravity (EG) agrees extremely well with experimental data. 
Nevertheless, it is not unreasonable to ponder other alternative or modified 
theories for gravity. Theoretically, the symmetry of general transformations 
of space-time coordinates, which dictates EG, is not guaranteed as the ultimate 
symmetry of space-time. Furthermore, while we believe that a theory of gravity 
should be a self-consistent quantum field theory, EG is not 
renormalizable\cite{chri}. On the cosmological side, we know that the big bang 
model, which is based on EG, has severe cosmological problems, despite its 
great success in modern cosmology. Although these problems can be circumvented 
by introducing an inflationary epoch in the early universe, a convincing 
microscopic origin for the slow-roll mechanism in new inflationary models is 
still lacking\cite{oliv}.

Among many modifications of EG, Brans-Dicke gravity (BDG) is the most popular 
one. Brans and Dicke, based on Mach's principle, introduced a scalar field 
$\hat{\phi}$ nonminimally coupled to the scalar curvature $R$, whose vacuum 
expectation value (vev) manifests as the effective gravitational constant 
$G$\cite{bran}. As a consequence, $G$ is no longer a physical constant and may 
change with time. This leads to models of so-called induced gravity in which 
an effective potential $V(\hat{\phi})$ is added to account for the evolution 
of $G$\cite{smol}. Recently, BDG was applied as a new approach to inflationary 
cosmology, designated the extended inflation, which attempted to give the old 
inflation a graceful exit\cite{la}. Later analyses showed that a successful 
extended inflation would require a non-trivial $V(\hat{\phi})$ or higher-order 
couplings of $\hat{\phi}$ to $R$\cite{acce}, thus rendering the approach more 
or less artificial.

In this {\em Letter}, we propose a new scalar-tensor theory of gravitation, 
based on scale invariance of space-time. We begin with the action,
\begin{equation}
  S_G= \int d^4 x \sqrt{g}\left[ -{1\over 2}\xi\hat{\phi}^2 R + 
    {1\over 2} g^{\mu\nu}\hat{\phi}_{;\mu}\hat{\phi}_{;\nu} 
    + {\lambda\over 4!}\hat{\phi}^4\right],
\label{actg}
\end{equation}
for pure gravity, with signature $(-,+,+,+)$ and $\xi=1/6$, where the 
semicolons denote taking covariant derivative.
The action is invariant under the scale transformations:
$g_{\mu\nu}\to \Omega^2(x)g_{\mu\nu}$, and 
$\hat{\phi}\to\Omega^{-1}(x)\hat{\phi}$. This is in fact the Weyl gravity 
(WG)\cite{weyl}. As is well known, WG is not physically acceptable because 
quantum phenomena provide an absolute standard of length. This can be resolved 
by either simply introducing a scale invariance breaking term such as a mass 
term for $\hat{\phi}$\cite{dese} or supposing that the Einstein equations 
refer to an space-time interval connecting two neighbouring points which is 
not the same as the interval measured by atomic apparatus\cite{dira}. Here we 
will follow the former wisdom, but instead formulate a scheme for breaking the 
scale invariance dynamically. This requires finding the quantum  corrections 
of the action $S_G$. Since $S_G$ has a fixed value of $\xi$ and a negative 
kinetic term for $\hat{\phi}$, it is generically different from BDG or induced 
gravity. The reader might worry that the model is tachyonic. As we will see 
below, a scrutiny shows that $\hat{\phi}$ is an auxiliary field and its 
one-loop effective potential can be well defined. As a consequence, the 
classical scalar field dynamics is similar to that in induced gravity, except 
that the kinetics of $\hat{\phi}$ contributes an effective negative 
stress-energy to the energy-momentum tensor. This peculiarity render the model 
a distinct feature that might shed a new light on understanding the inflation 
physics. Other issues, such as metric perturbations, varying gravitational 
constant, and the cosmological constant will also be briefly discussed.

If we were living in a pure-gravity space-time, the physical laws in it would 
be well described by WG. In fact, the pure WG~(\ref{actg}) is equivalent to 
the EG plus a cosmological constant. This can be seen by simply choosing the 
Einstein gauge: $\hat{\phi}^2=(3/4\pi)M_P^2$, where $M_P$ is Planck mass. 
Thus, we can treat WG as a generalization of EG that admits space-time varying 
gravitational `constant' and cosmological `constant'. To prove this rigorously, 
we perturb the Lagrangian~(\ref{actg}) around the ground-state background:
\begin{equation}
g_{\mu\nu}=\eta_{\mu\nu} + \phi^{-1} h_{\mu\nu},\quad \hat{\phi}=\phi +\sigma,
\label{expand}
\end{equation} 
where $\eta_{\mu\nu}$ is a flat space-time, and $\phi$ is a constant background 
field.
Redefining $\rho_{\mu\nu} = h_{\mu\nu} +2\sigma\eta_{\mu\nu}$, we find that 
the spectrum consists of a graviton $\rho_{\mu\nu}$ with a mass induced by the 
non-zero vacuum energy, as well as an auxiliary field $\sigma$ that can be 
eliminated at the quadratic level. This spectrum has no difference from the 
EG's. We thus establish the first result: the Weyl scalar field, unlike 
Brans-Dicke's or other induced-gravity models', is a non-dynamical
degree of freedom. 

Now we evaluate the one-loop effective potential for $\phi$ by using the 
background field method\cite{abbo}. After the expansion~(\ref{expand}) up to 
terms quadratic in the quantum fields $h_{\mu\nu}$  and $\sigma$, we add the 
gauge fixing term,
\begin{equation}
{\cal L}_{\rm gf} = {1\over 2\alpha}                                      
(\partial^{\mu}h_{\mu\rho} - {1\over 2} \partial_{\rho}h_{\mu}^{\mu})^2
 + {1\over 2\beta} (\partial_{\mu}\sigma)^2,
\end{equation}
where $\alpha$ and $\beta$ are arbitrary parameters. In the one-loop level, 
the ghost fields do not couple with other fields and thus can be neglected. 
Integrating out the quantum fields, we obtain the unrenormalized one-loop 
effective potential
\begin{equation}
  V(\phi) = {\lambda\phi^4\over 4!} +
    3{\rm Tr}\ln\left( q^2+{1\over 2}\lambda\phi^2\right) + f(\alpha,\beta),
\end{equation}
where $q$ is a four-momentum, and the last term is gauge-choice dependent with 
$f(0,0)=0$. Henceforth, we choose the Laudau-DeWitt gauge,
$\alpha=\beta=0$\cite{hugg}. Then, the standard renormalization 
procedure\cite{cole} gives the renormalized effective potential,
\begin{equation}
  V(\phi) = {\kappa\lambda^2\phi^4\over 64\pi^2}\left( 
    2\ln \frac{\phi}{v} -{1\over 2}\right) + \Lambda,
\label{pot}
\end{equation}
where $\kappa=3/2$\cite{com1}, and $v={\sqrt {3/4\pi}}M_P$. Also, we have 
chosen the vev $\langle\phi\rangle=v$ and 
$\Lambda=\kappa\lambda^2 v^4/(128\pi^2)$ such that $V(v)=0$ and its first 
derivative $V'(v)=0$. Apparently, this potential breaks the scale-invariance 
symmetry, and the vev determines the ground-state gravitational constant 
$G_N=3/(4\pi v^2)$.

Since $\phi$ originates from the scaling of space-time, we do not expect that 
it couples directly to classical matter. Thus, we write down the action of 
the universe as $S_U=\tilde S_G+S_M$, where $\tilde S_G$ is the effective 
gravity action given by (\ref{actg}) except replacing the quartic potential 
with $V(\phi)$, and $S_M$ is the action for classical matter.
By varying $S_U$ with respect to $g^{\mu\nu}$ and $\phi$ respectively, we 
obtain the field equations,
\begin{eqnarray}
&&{1\over 6}\phi^2 \left( R^{\mu\nu}-{1\over 2} R g^{\mu\nu}\right) =   
-\left(T_\phi^{\mu\nu}+T_M^{\mu\nu}\right),
\label{feq1} \\ 
&&T_\phi^{\mu\nu}=-\phi^{;\mu}\phi^{;\nu}+{1\over 2}\phi^{;\rho}\phi_{;\rho}
g^{\mu\nu} -{1\over 6} (\phi^2)^{;\rho}{}_{;\rho} g^{\mu\nu} \nonumber \\
&&\quad\quad\quad +{1\over 6}(\phi^2)^{;\mu;\nu}+g^{\mu\nu}V(\phi), 
\label{feq2} \\
&&\phi^{;\mu}{}_{;\mu}+{1\over 6}\phi R -V'(\phi)=0,
\label{feq3}
\end{eqnarray}
where $R^{\mu\nu}$ is the Ricci tensor, and $T_M^{\mu\nu}$ is the matter 
stress-energy defined by $2\delta S_M=\int d^4x {\sqrt g} T_M^{\mu\nu}\delta 
g_{\mu\nu}$.
Note that $T_\phi^{\mu\nu}$ is negative. It can be shown from the field 
equations that
\begin{eqnarray} 
T_M^{\mu\nu}{}_{;\nu} &=& 0, \label{coe} \\
-g_{\mu\nu} T_M^{\mu\nu} &=& 4V(\phi)-\phi V'(\phi). \label{trace}
\end{eqnarray}

Let us assume Robertson-Walker space-time and a perfect-fluid form for matter:
\begin{eqnarray}
d\tau^2 &=& dt^2-a^2(t) \left[{dr^2\over 1-kr^2}+r^2 d\Omega^2 \right], \\
T_M^{\mu\nu} &=& p g^{\mu\nu} + (\rho+p)U^\mu U^\nu,
\end{eqnarray}
where $a(t)$ is the cosmic scale factor, $\rho$ and $p$ are respectively the 
matter energy density and pressure, and $U^\mu$ is the four-velocity with 
$U^\mu U_\mu=-1$. 
Then, from the field equations and Eqs.~(\ref{pot}), (\ref{coe}) and 
(\ref{trace}), we obtain the following evolution equations,
\begin{eqnarray}
&&{\dot a^2\over a^2}+2{\dot a\over a}{\dot {\phi}\over \phi}={2\over
  \phi^2}\left[-{\dot {\phi}^2\over 2}-V(\phi)+\rho\right]-{k\over a^2},
\label{eeq1} \\
&&\ddot{\phi}+3{\dot a\over a}\dot {\phi} +\left({\ddot a\over a}+{\dot
  {a}^2\over a^2}+{k\over a^2}\right)\phi+V'(\phi)=0, 
\label{eeq2} \\
&&{dp\over dt}a^3={d\over dt}\left[a^3(\rho+p)\right], 
\label{eeq3} \\
&&\rho-3 p=4V(\phi)-\phi V'(\phi)=4\Lambda\left(1-\frac{\phi^4}{v^4}\right),
\label{eeq4}
\end{eqnarray}
where the dot represents time differentiation.

The evolution of $\phi$ is a damped motion in an effective potential 
$U(\phi)=-\phi^2 R/12+V(\phi)$. $U(\phi)$ will have a maximum at $\phi_1$ and 
a global minimum at $\phi_0$ such that $\phi_1<v/{\sqrt e}<\phi_0<v$ if 
\begin{equation}
-R<3\kappa\lambda^2 v^2/(16\pi^2 e^{1\over 2}).
\label{bound}
\end{equation}
Suppose initially $\phi$ is located at the local minimum $\phi=0$. Then, 
Eqs.~(\ref{eeq1}) and (\ref{eeq4}) imply that $\rho=-p=\Lambda$. Usually, this 
equation of state signifies an exponential growth of $a$. But, in the strong 
gravitational coupling limit (i.e., $\phi<<v$) such as here, 
$H\equiv \dot a/a$ is arbitrary. As the equation of state is homogeneous in 
space-time, we assume the metric be the maximally symmetric de Sitter space 
with $R=-12{\bar H}^2$, where $\bar H$ is a constant parameter only subject 
to the constraint~(\ref{bound}). As such, $\phi$ will make a quautum tunneling 
through the barrier, $\Delta U=U(\phi_1)-U(0)$, via the Hawking-Moss (HM) 
homogeneous bubble solution\cite{haw1,mot1,jens,deru,lind}: the universe 
undergoes a quantum transition everywhere to $\phi=\phi_1$. It is interpreted 
as that the bubble nucleation occurs in a spatially homogeneous manner over a 
region of the order of the event horizon ${\bar H}^{-1}$, from which the whole 
of the presently observed universe would have developed\cite{haw2}. This has 
been criticized by several authors\cite{jens}. However, our point of view is 
that one cannot jump into any conclusion until the ambiguity of interpretation 
of the results of Euclidean approach to this tunneling problem has been 
clarified. (See also Ref.~\cite{lind}.) In the following, we will adopt HM's 
interpretation and explore the consequences to the present context.
The tunneling probability per unit four-volume is of the order of ${\bar H}^4 
e^{-B}$, where 
\begin{equation}
B=\frac{8\pi^2}{3}\frac{\Delta U}{{\bar H}^4}
=\frac{64\pi^4}{3\kappa\lambda^2}\ln^{-1}\frac{v}{\phi_1}
\left( 1-{1\over 4} \ln^{-1}\frac{v}{\phi_1} \right).
\end{equation}
Thus, a small $\lambda$ would induce a huge tunneling action, and the causal 
region could easily inflate so large as to solve the cosmological problems.

After inflation, $\phi$ will roll quickly down the potential with critical 
damping towards $\phi_0$. During the radiation-dominated (RD) era, the trace 
condition~(\ref{eeq4}) simply reads $\phi=v$. Thus, the evolution of the 
universe has no difference from the standard big-bang. In the matter-dominated 
(MD) epoch, the evolution equations can be rewritten as
\begin{eqnarray}
&&\left(1-\frac{1}{2H}\frac{\dot G}{G}\right)^2 = \Omega_M-\Omega_\phi, \\
&&1-\frac{\phi^4}{v^4} = \frac{\rho_0}{4\Lambda}\left(\frac{a_0}{a}\right)^3,\\
&&\frac{\dot G}{G}={3\over 2} \left(1-\frac{v^4}{\phi^4}\right)H,
\end{eqnarray}
where the curvature term has been neglected, $\rho_0$ is the present energy 
density, and we have defined
\begin{equation}
G\equiv \frac{3}{4\pi\phi^2},\quad \Omega_M\equiv \frac{8\pi G\rho}{3H^2},
\quad
\Omega_\phi\equiv \frac{8\pi G V(\phi)}{3H^2}.
\end{equation}
>From this we deduce that the discrepancy from the standard big-bang is 
measured by the factor $\rho_0/\Lambda \simeq 
10^{-118}\lambda^{-2}{\Omega_M}_0 h_0^2$, which would be almost a zero unless 
$\lambda$ is fine-tuned. Below we will show that galaxy formation would 
require $\lambda$ to be about $10^{-2}$. Thus, after inflation $G$ is 
practically equal to the gravitational constant $G_N$. Also, if we identify 
$\Omega_\phi$ as the cosmological constant, then the latter would be almost 
vanishing.

Now we turn to estimate the metric perturbations generated during the HM 
bubble nucleation. For scalar perturbation, one needs to calculate the 
quantity $(\delta \rho)_k/(\rho+p)$ at the time the Fourier mode with 
wavenumber $k$ crosses outside the horizon during inflation\cite{bard}. 
>From the evolution equations, we find that $\rho+p\sim \dot\phi^2$ for 
$\phi<\phi_1$. On the other hand, the HM solution reflects the fact that the 
most probable quantum fluctuation will only just get over the potential 
barrier\cite{haw2}. It means that the de Sitter thermal fluctuation of the 
field represented by $\dot\phi^2$ should be approximately equal to $\Delta U$. 
(See also Ref.~\cite{mot2}.) To find $(\delta \rho)_k$, we compute the 
spectral energy density of quanta produced during the transition of $\phi$ by 
solving the equation of motion for its fluctuation $\delta\phi(t,\vec x)$,
\begin{equation}
\ddot{\delta\phi} + 3{\bar H}\dot{\delta\phi} 
+ U''(\phi)\delta\phi - a^{-2}\nabla^2\delta\phi=0.
\end{equation}
We approximate the transition as a sudden process: $U''(\phi)$ is equal to 
$U''(0)$ and $U''(\phi_1)$ respectively before and after the transition. 
Following the standard Bogoliubov transformation\cite{birr}, we find that 
after the transition the $(\delta \rho)_k$ at horizon crossing is 
approximately given by ${\bar H}^4$ for cosmologically interesting scales. 
Hence, when the mode crosses back inside the horizon during the MD era, the 
scalar perturbation is 
\begin{equation}
\left.\frac{(\delta \rho)_k}{\rho}\right\vert_{\rm hor}
\sim \frac{{\bar H}^4}{\Delta U} = \frac{8\pi^2}{3B}.
\label{sp}
\end{equation}
This scale-invariant result is similar to that given in Ref.~\cite{mot2}.
As is well known, in standard Einstein gravity the amplitude of tensor 
perturbation behaves as a massless, minimally coupled scalar field, and their 
Fourier modes are related by $h_k={\sqrt {16\pi G_N}}\psi_k$\cite{gris}. As 
such, the production of tensor mode $h_k$ in inflationary cosmology can be 
computed from the quantum fluctuation of $\psi_k$ during inflation and its 
consequent evolution. The result for an exponential inflation is a 
scale-invariant spectrum with amplitude ${\sqrt {G_N}}{\bar H}$\cite{ruba}. 
This analysis could be directly applicable here only if we replace $G_N$ by 
$G$. Hence, when a tensor mode re-enters the horizon during the MD era, $G$ 
equals to $G_N$ and the mode amplitude is  
\begin{equation}
\left.h_k\right\vert_{\rm hor} \sim G_N^{1\over 2}{\bar H}.
\label{tp}
\end {equation}
Here the novelty is that the scalar and tensor perturbations depend 
respectively on two unrelated parameters, $\lambda$ and ${\bar H}$. Thus, 
they would independently induce temperature fluctuations in the cosmic 
microwave background (CMB). From the recent $COBE$ measurement of the CMB 
large-scale anisotropy, $\Delta T/T\sim 10^{-5}$\cite{smoo}, Eqs.~(\ref{sp}) 
and (\ref{tp}) imply that
\begin{equation}
\lambda\;^<_\sim\;10^{-2}{\sqrt c},\quad{\rm and}\quad {\bar H}/M_P
\;^<_\sim \;10^{-5},
\end{equation}
where $c$ is a logarithmically dependent factor. While almost all inflation 
models have negligible tensor contribution for a scale-invariant 
spectrum\cite{stei}, the tensor mode produced here can easily dominate the 
scalar mode by suitably tuning $\lambda$ and ${\bar H}$. In particular, it 
was recently pointed out that the CMB large-scale anisotropy could be due in 
part to the tensor perturbation\cite{krau}. 

In conclusion, we have attempted to include the scale-invariance as a 
fundamental symmetry of space-time. This results in a new scalar-tensor theory 
of gravitation. It can be realized as a Brans-Dicke model with a negative BD 
parameter. Our model has a very interesting cosmology. Firstly, it naturally 
incorporates a modified Hawking-Moss homogeneous inflation in which the Hubble 
expansion parameter is not fixed by the vacuum energy. While the interpretation 
of the HM solution is not without arbitrariness, we proceeded to estimate the 
metric perturbations produced during the HM inflation. We have found that 
both the scalar and tensor perturbations have scale-invariant spectra but 
their amplitudes are independent of each other. In light of this, the CMB 
large-scale anisotropy might be mainly induced by the tensor mode. Secondly, 
if both the time variation of the gravitational constant and the non-vanishing 
of the cosmological constant are ascribed to a single theory of gravitation, 
our model would suggest that
they are in fact exceedingly small after inflation. Here we have provided 
neither a detailed picture of how the universe changes from one phase to 
another, nor a systematic method of calculating the metric perturbations 
generated during the HM inflation. The work about these is in 
progress. 

We would like to thank H. T. Cho, H.-C. Kao, W. F. Kao, and S.-Y. Wang for 
their useful discussions. 
This work was supported in part by the National Science Council, 
ROC under the Grants NSC86-2112-M-001-009 and
NSC87-2112-M-001-039.


\begin{references}
\bibitem{com1}
We have also used the gauge-choice independent Vilkovisky-DeWitt 
method\cite{vilk,hugg} to calculate the effective potential. The obtained 
value for $\kappa$ is only slightly different from $3/2$\cite{ling}.
\end{references}
\end{document}